%% file: main.tex
\documentclass[10pt,journal,compsoc]{IEEEtran}
\usepackage[hyphens]{url}  % DO NOT CHANGE THIS
\usepackage{graphicx} % DO NOT CHANGE THIS

\usepackage{amsfonts} 
\usepackage{subfigure}
\usepackage{multirow}

\usepackage{enumitem}
\usepackage{amsmath}
\usepackage{amsthm}

\ifCLASSOPTIONcompsoc
  \usepackage[nocompress]{cite}
\else
  \usepackage{cite}
\fi

\ifCLASSINFOpdf
\else
\fi

\begin{document}
\title{GDSRec: Graph-Based Decentralized Collaborative Filtering for Social Recommendation}

\author{Jiajia~Chen,
        Xin~Xin,
        Xianfeng~Liang,
        Xiangnan~He,
        and~Jun~Liu,~\IEEEmembership{Senior~Member,~IEEE}%
\IEEEcompsocitemizethanks{\IEEEcompsocthanksitem J. Chen, X. He, J. Liu are with the School of Information Science and Technology, University of Science and Technology of China, Hefei, Anhui 230026, China.\protect\\
E-mail: jia2chan@mail.ustc.edu.cn, \{hexn, junliu\}@ustc.edu.cn
\IEEEcompsocthanksitem X. Xin is with the School of Computer Science and Technology, Shandong University, Qingdao, Shandong 266237, China.\protect\\
E-mail: xinxin@sdu.edu.cn% <-this % stops an unwanted space
\IEEEcompsocthanksitem X. Liang is with the School of Data Science, University of Science and Technology of China, Hefei, Anhui 230026, China.\protect\\
E-mail: zeroxf@mail.ustc.edu.cn}% <-this % stops an unwanted space
}

\markboth{Journal of \LaTeX\ Class Files,~Vol.~14, No.~8, August~2015}%
{Shell \MakeLowercase{\textit{et al.}}: Bare Demo of IEEEtran.cls for Computer Society Journals}

\IEEEtitleabstractindextext{%
\begin{abstract}
Generating recommendations based on user-item interactions and user-user social relations is a common use case in web-based systems. These connections can be naturally represented as graph-structured data and thus utilizing graph neural networks (GNNs) for social recommendation has become a promising research direction. However, existing graph-based methods fails to consider the bias offsets of users (items).
For example, a low rating from a fastidious user may not imply a negative attitude toward this item because the user tends to assign low ratings in common cases.
Such statistics should be considered into the graph modeling procedure. While some past work considers the biases, we argue that these proposed methods only treat them as scalars and can not capture the complete bias information hidden in data. Besides, social connections between users should also be differentiable so that users with similar item preference would have more influence on each other. To this end, we propose Graph-Based Decentralized Collaborative Filtering for Social Recommendation (GDSRec). GDSRec treats the biases as vectors and fuses them into the process of learning user and item representations. The statistical bias offsets are captured by decentralized neighborhood aggregation while the social connection strength is defined according to the preference similarity and then incorporated into the model design. We conduct extensive experiments on two benchmark datasets to verify the effectiveness of the proposed model. Experimental results show that the proposed GDSRec achieves superior performance compared with state-of-the-art related baselines. Our implementations are available in \url{https://github.com/MEICRS/GDSRec}.
\end{abstract}

}

\maketitle

\IEEEdisplaynontitleabstractindextext

% \fi
\IEEEpeerreviewmaketitle

\input{introduction}

\input{background}
\input{methodology}

\input{experiments}
\input{conclusions}

\appendices

\ifCLASSOPTIONcompsoc
  \section*{Acknowledgments}
\else
  \section*{Acknowledgment}
\fi

This work is supported by the National Natural Science Foundation of China (U19A2079, 62121002), the Youth Innovation Promotion Association CAS (CX2100060053), and USTC Tang Scholar. Xin Xin and Jun Liu are the corresponding authors.

\ifCLASSOPTIONcaptionsoff
  \newpage
\fi




\bibliographystyle{IEEEtran}
\linespread{1.0}\selectfont 
\bibliography{reference}


\vspace{-2cm}
\begin{IEEEbiography}[{\includegraphics[width=1in,height=1.25in,clip,keepaspectratio]{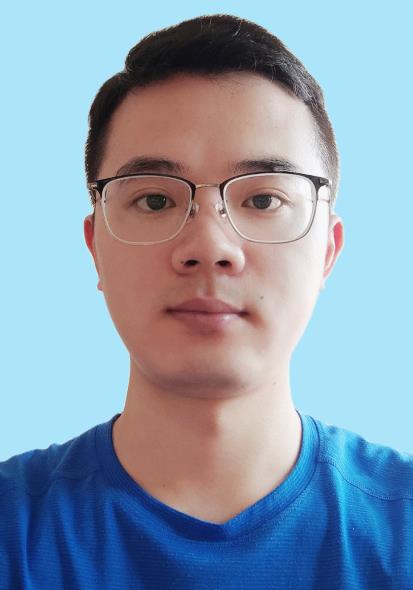}}]{Jiajia Chen} is currently working towards the Ph.D. degree at University of Science and Technology of China (USTC), Hefei, China. His research interests include data mining and recommender systems.
\end{IEEEbiography}

\vspace{-2cm}
\begin{IEEEbiography}[{\includegraphics[width=1in,height=1.25in,clip,keepaspectratio]{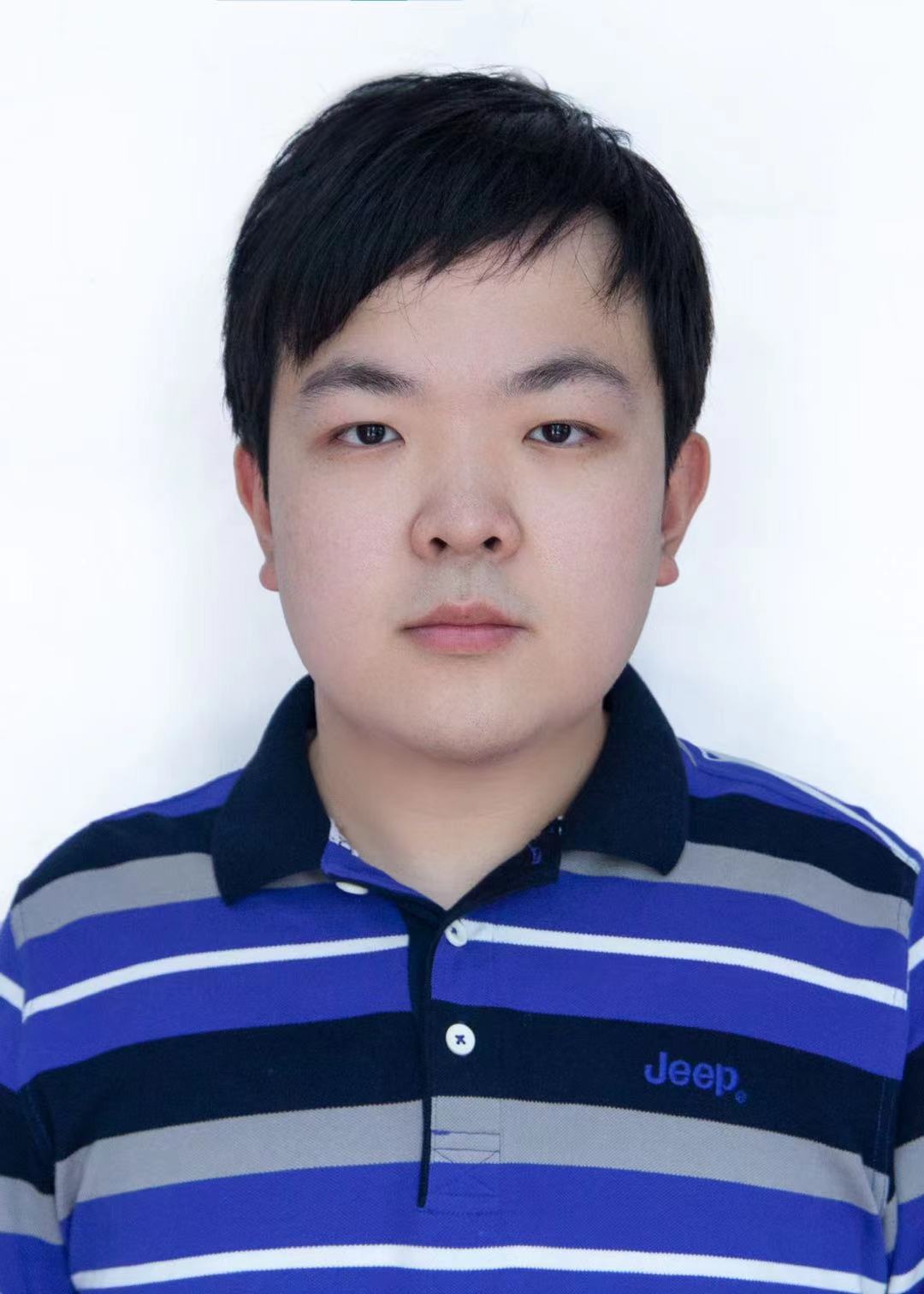}}]{Xin Xin} is now a tenure-track assistant professor in the school of computer science and technology, Shandong University. His research interests include machining learning and reinforcement learning for recommender systems and information retrieval. He has published more than 20 papers in top-ranking conferences, including SIGIR, IJCAI, ACL, WSDM, ect. He also serves as the program committee member and invited reviewer for tire-1 con- ferences and journals, such as SIGIR, IJCAI, WSDM, ACL, ACM Multimedia, TKDE, TOIS.
\end{IEEEbiography}

\vspace{-2cm}
\begin{IEEEbiography}[{\includegraphics[width=1in,height=1.25in,clip,keepaspectratio]{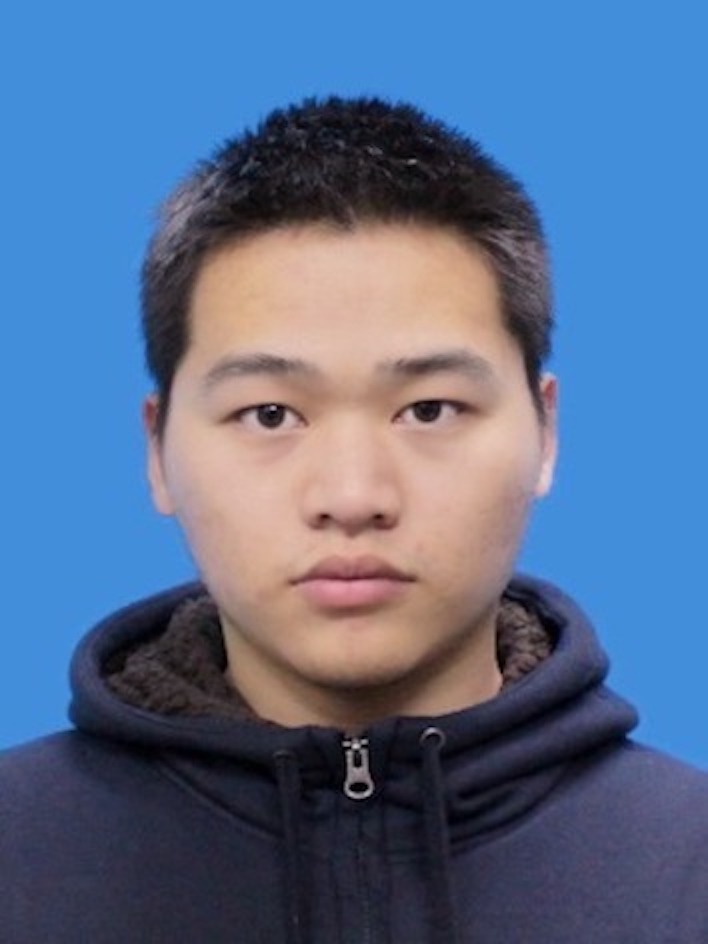}}]{Xianfeng Liang} received the M.S. degree from University of Science and Technology of China (USTC), Hefei, China, in 2021. His major research interests include data mining, machine learning and optimization.
\end{IEEEbiography}

\vspace{-2cm}
\begin{IEEEbiography}[{\includegraphics[width=1in,height=1.25in,clip,keepaspectratio]{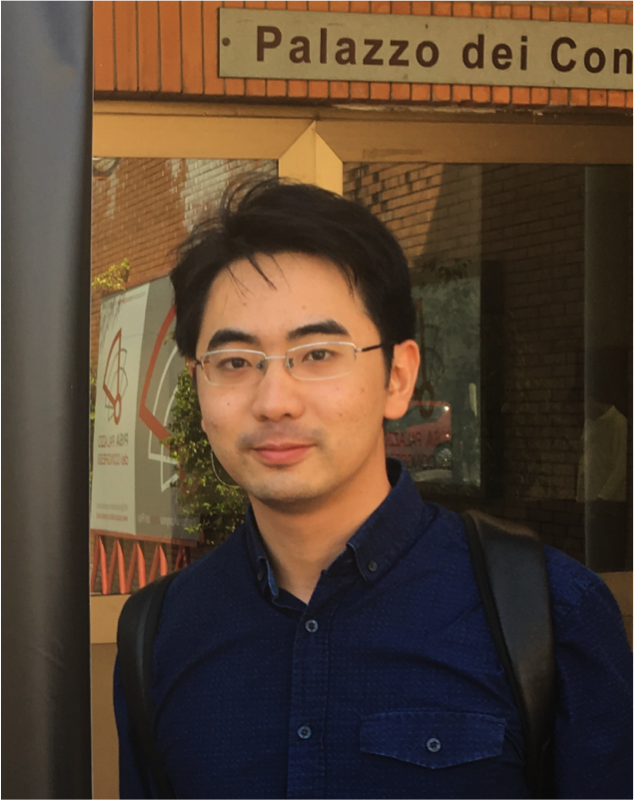}}]{Xiangnan He} is a professor at the University of Science and Technology of China (USTC). His research interests span information retrieval, recommendation, data mining, and multimedia. He has over 100 publications that appeared in top conferences such as SIGIR, WWW, and KDD, and journals including TKDE, TOIS, and TNNLS. His work has received the Best Paper Award Honorable Mention in SIGIR (2021, 2016), and WWW 2018. He is serving as the associate editor for ACM Transactions on Information Systems (TOIS), IEEE Transactions on Big Data (TBD), and the SPC/PC member for top conferences including SIGIR, WWW, KDD, MM, WSDM, ICML etc.
\end{IEEEbiography}

\vspace{-2cm}
\begin{IEEEbiography}[{\includegraphics[width=1in,height=1.25in,clip,keepaspectratio]{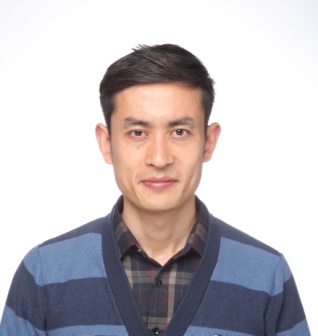}}]{Jun Liu} (S’11–M’13–SM’16)  received the B.S. degree in mathematics from Wuhan University of Technology, Wuhan, China, in 2006, the M.S. degree in mathematics from Chinese Academy of Sciences, China, in 2009, and the Ph.D. degree in electrical engineering from Xidian University, Xi'an, China, in 2012.
From July 2012 to December 2012, he was a Post-doctoral Research Associate with the Department of Electrical and Computer Engineering, Duke University, Durham, NC, USA. From January 2013 to September 2014, he was a Postdoctoral Research Associate with the Department of Electrical and Computer Engineering, Stevens Institute of Technology, Hoboken, NJ, USA. He is currently an Associate Professor with the Department of Electronic Engineering and Information Science, University of Science and Technology of China, Hefei, China. His research interests include statistical signal processing, image processing, and machine learning. 
Dr. Liu is a Member of the Sensor Array and Multichannel (SAM) Technical Committee, IEEE Signal Processing Society. He is the co-author of a book Advances in Adaptive Radar Detection and Range Estimation (Springer, 2022). He was the recipient of the Best Paper Award from the IEEE WCSP 2021. He is currently an Associate Editor for the IEEE Signal Processing Letters, and a Member of the Editorial Board of the Signal Processing (Elsevier). 
\end{IEEEbiography}

\end{document}

%% file: introduction.tex
\IEEEraisesectionheading{\section{Introduction}
\label{sec1}}
%%%%%%%%%%%%%%%%%%  %%%%%%%%%%%%%%%%%%%%%
\IEEEPARstart{I}{n} the face of a huge number of web users and information explosion, recommender systems are of vital importance which can alleviate information overload and provide users with more efficient and high-quality services. An effective recommender system can benefit both users by acquiring their preferred contents (e.g. movies, music, merchandise) from a large amount of information, and service providers by reducing promotional costs. As a result, recommender systems have attracted widespread interests in recent years. 
%A variety of Internet companies, such as Alibaba, Amazon, Facebook, etc., and many academics are pushing for this technology. 
%Generally speaking, there are two main types of recommender algorithms: content-based algorithms \cite{PazzaniBillsus07} and collaborative filtering algorithms \cite{GoldbergNichols92}. Among these methods, collaborative filtering has been widely used because of its effectiveness and scalability. Matrix factorization \cite{HuKoren08,KorenBell09}, one of the most important collaborative filtering methods, has achieved great success in various scenarios \cite{LiuRogers17,SmithLinden17}. It maps users and items to a shared latent factor space, and interactions between users and items are modeled by the inner product of their latent factors.
Meanwhile, exploiting social relations to improve the performance of recommendation has also become increasingly popular with the growth of social media \cite{MaZhou11,TangHu13a,TangWang16}.  %\cite{MaZhou11,TangHu13a,TangWang16} demonstrated that exploiting because social relations can contribute to improve recommendation performance
In social networks, there is a flow of information among connected friends. A user's preference is similar to or influenced by the people around him/her, which has been proved by social correlation theories \cite{MarsdenFriedkin93,McPhersonSmith01}. 

%In \cite{WengLimwsdm10}, the authors found that users with following relations are more likely to share similar interests in topics than two randomly chosen users. Moreover, users with strong ties are more likely to share similar tastes than those with weak ties, treating all social relations equally is likely to lead to degradation in recommendation performance \cite{TangHu13b}. Therefore, the social relations play a significant role to help the users filter information. Many social-based recommendation algorithms have been presented to improve the recommendation accuracy \cite{MaYang08,JamaliEster10,MaZhou11,YangLei16}. In \cite{MaYang08}, a factorization approach was proposed to solve the rating prediction problem by employing both the users' social network information and rating records. TrustMF \cite{YangLei16} was introduced by using the matrix factorization of the trust network between the users in social interactions for the rating prediction. One can find comprehensive overviews about social recommendations in \cite{TangHu13b}.

Recently, deep learning has shown strong capability to achieve good performance due to its high expressiveness and model fidelity. Graph neural networks (GNNs) utilize the advances of deep learning for graph-structured data and have been applied for various fields such as geo-location \cite{nassar2020geograph, rahimi2018semi} and bio-informatic \cite{fout2017protein,liu2020deep}.
% Despite utilizing deep learning for feature extraction (e.g.,music audio features \cite{VanDieleman13} and image visual content \cite{ZhaoLu16}), attempts have also been made to directly use deep learning to model user-item interactions. 
For the domain of social recommendation, the user-item interactions and user-user friendship can be naturally represented as graphs, in which users and items are the nodes while interactions (friendship) are the edges. Based on such observations, utilizing the recent advance of GNNs for social recommendation has become a promising research direction.

However, almost GNN-based recommendation methods are learned from the original interaction graph with little attention paid to the statistical information of the graphs, which could result into misunderstanding of the real user preference. For example, from the user perspective, a fastidious user may tend to give a low rating to every movie he has watched, then a rating of 3 out of 5 may actually denote a positive preference of this user. From the item perspective, a rating 4 of 5 could also represent a negative attitude if the average rating for this item can achieve a high score (e.g., 4.5). This is a bias hidden in the data that would mislead the training for users and items representations. The author in \cite{koren2008factorization} has introduced similar considerations. Based on this insight, FunkSVD and SVD++ have been proposed, which model these user and item biases as scalars in rating predictions. However, we argue that these methods are simple but do not capture the true bias hidden in practice.  Firstly, we consider that using scalar is not enough to completely model the real biases of users and items. Secondly, existing methods do not explicitly construct the biases in data, but use original data to learn. It also lead to an inability to learn the bias well.
Besides, the social connection strength should also be differentiable. In  \cite{TangHu13b}, it demonstrates that users with strong connections are more likely to share similar tastes than those with weak connections and thus treating all social relations equally would also lead to sub-optimal solutions.

In this paper, we design a new GNN-based model to address the above problems for social recommendation. More precisely, we treat the biases as vectors and fuse them into user/item representations in the proposed model. This design could help us to learn the representations well. To this end, we design a decentralized interaction graph to consider the statistical bias offsets of users (items). This graph is constructed by extracting bias information explicitly and helps the model learn better representations. Besides, we re-weight the user-user connections according to the preference similarity, which can help the model focus on useful friendship connections while denoise the redundant aggregation.
Our major contributions are summarized as follows:
\begin{itemize}[leftmargin=*]
	\item We treat the rating biases as vectors and fuse them into the process of learning user and item representations. To this end, we introduce a new perspective to process the original graph to a decentralized graph and learn the user and item representations from it. The decentralized graph is acquired through exploiting the statistical information of the original data, thus the bias information is extracted explicitly on the graph.
	\item A simple yet effective explicit strength of the social connection is given, which can be then incorporated into the final prediction rule and helps to improve the recommendation performance.
	%The explicit strength of the social relation can help to improve the prediction accuracy.
	\item We propose a new GNN-based collaborative filtering model (GDSRec) for social recommendation, which are learned on the decentralized graph with explicit differentiable social connection strengths.
	\item We conduct experiments on two real-world datasets to verify the effectiveness of the proposed model. Experimental results show that GDSRec outperforms the compared state-of-the-art baselines.
\end{itemize}

The remainder of this paper is organized as follows. In Section \ref{sec2}, we introduce the background including related works and some notations. The proposed framework is detailed in Section \ref{sec3}. In Section \ref{sec4}, experiments on two real-world datasets are conducted to validate the proposed approach. Finally, we give a conclusion of this paper in Section \ref{sec5}.

%% file: background.tex
\section{Background}
\label{sec2}
\subsection{Related Work}
For the general recommendation task, there are two main types of algorithms: content-based algorithms \cite{PazzaniBillsus07} and collaborative filtering algorithms \cite{GoldbergNichols92, JinSi03}. Among these methods, collaborative filtering has been widely used because of its effectiveness and scalability. Matrix factorization \cite{HuKoren08,KorenBell09,GuZhou10}, one of the most important collaborative filtering methods, has achieved great success in various scenarios \cite{LiuRogers17,SmithLinden17}. It maps users and items to a shared latent factor space, and interactions between users and items are modeled by the inner product of their latent factors. In \cite{koren2008factorization}, SVD++ considers user and item biases that extends matrix factorization model. Besides, explorations of social networks for recommendation have been proved to be effective. In \cite{WengLimwsdm10}, the authors found that users with following relations are more likely to share similar interests in topics than two randomly chosen users. Therefore, the social relations play a significant role to help the users filter information. Based on such observation, many social-based recommendation algorithms have been presented \cite{MaYang08,JamaliEster10,MaZhou11,YangLei16,WangLu16,WangHoi17}. In \cite{MaYang08}, a factorization approach was proposed to solve the rating prediction problem by employing both the users' social network information and rating records. TrustMF \cite{YangLei16} was introduced by using the matrix factorization of the trust network between the users in social interactions for the rating prediction. TrustSVD \cite{guo2015trustsvd} exploits trust information that extends SVD++. The authors in \cite{WangLu16,WangHoi17} utilized Jaccard's coefficient to compute the strength of the social relations, but did not take the ratings into account. Furthermore, some other works utilized side information to improve the recommendation quality, e.g. \cite{YuRen14,ShiHu18}. Recently, SREPS \cite{liu2018social} learns the user's multiple preference in different scenarios. 

Deep learning models have also been exploited to enhance the model expressiveness for recommendation.
Due to the fact that the recommendation data can be naturally organised as graphs, research about exploiting GNNs \cite{ThomasWelling16,DefferrardBresson16,HamiltonYing17,MaWang19} for recommendation have also been conducted. The key insight of GNNs is to learn the representations of the nodes by aggregating feature information from neighborhoods. This conforms the nature of collaborative filtering. 
In \cite{BergKipf17}, the authors proposed a graph auto-encoder framework called graph convolution matrix completion (GCMC) based on differentiable message passing on the user-item interaction graph without using the social relations. Neural graph collaborative filtering (NGCF) \cite{WangHesigir19} was proposed to integrate the user-item interactions into the embedding
process.Knowledge graph attention network for recommendation (KGAT) \cite{WangHesigkdd19} was introduced to explicitly model the high-order connectivities in knowledge graph. DiffNet \cite{WuSun19} was proposed by using a layer-wise influence diffusion part to model how users' latent preferences are recursively influenced by trusted users. A GNN-based social recommendation algorithm, GraphRec, was introduced in \cite{FanMa19}. It provides an approach to jointly capture the interactions and the ratings for learning the representations of the users and the items. Further, the social network information is employed in GraphRec for learning better representations of the users. DANSER \cite{wu2019dual} proposes two dual graph attention networks to learn deep representations for social effects in recommender systems. Diffnet++ \cite{wu2020diffnet++} promotes the user and item representations by injecting both the higher-order user latent interest reflected in the user-item graph and higher-order user influence reflected in the user-user graph. LightGCN \cite{he2020lightgcn} simplifies NGCF with removing nonlinear activation and feature transformation in graph convolution networks and promotes the performance of recommendation. ESRF \cite{yu2020enhance} develops a deep adversarial framework based on graph convolution networks to address the challenges of social recommendation. FBNE \cite{chen2020social} explores the implicit higher-order user-user relations though folding a user-item bipartite graph to improve the performance of social recommendation. HOSR \cite{liu2020modelling} is to generate user embedding by performing embedding propagation along high-order social neighbors. However, most of these methods fail to consider the statistic offsets existing in the graph data while also lack an effective yet simple design for the strength of social relations. 

\subsection{Notations}
Let $\mathcal{O}$ be the set of observed ratings (i.e., $r_{ij} \neq 0$). $R(v_j)$ is the set of users who have interacted with the item $v_j$ and $R(u_i)$ is the set of items which the user $u_i$ has interacted with. Let $N(u_i)$ be the set of users whom the user $u_i$ connects within the social network directly. The vector $\mathbf{p}_{u_i} \in \mathbb{R}^D$ denotes the embedding of user $u_i$, and $\mathbf{q}_{v_j} \in \mathbb{R}^D$ represents the embedding of item $v_j$, where $D$ is the embedding size. $E(u_i)$ and $E(v_j)$ denote the average ratings of the user $u_i$ and the item $v_j$, respectively. $\lceil\cdot\rceil$ and $|\cdot|$ are the integer-valued function and the absolute-valued function, respectively. $\oplus$ denotes the concatenation operation between two vectors. $\langle\cdot\rangle$ is used to obtain the number of entries in a dataset.

%% file: methodology.tex
\section{The Proposed Framework}
\label{sec3}
We first give the problem formulation, then introduce the decentralized graph and the proposed framework. Later, we detail how to get a predicted rating and learn the user/item latent factor offsets from the proposed model. Finally, we explain how to train the model.

\begin{figure}
	\centering
	\setlength{\abovecaptionskip}{0cm}
	\includegraphics[width=0.48\textwidth]{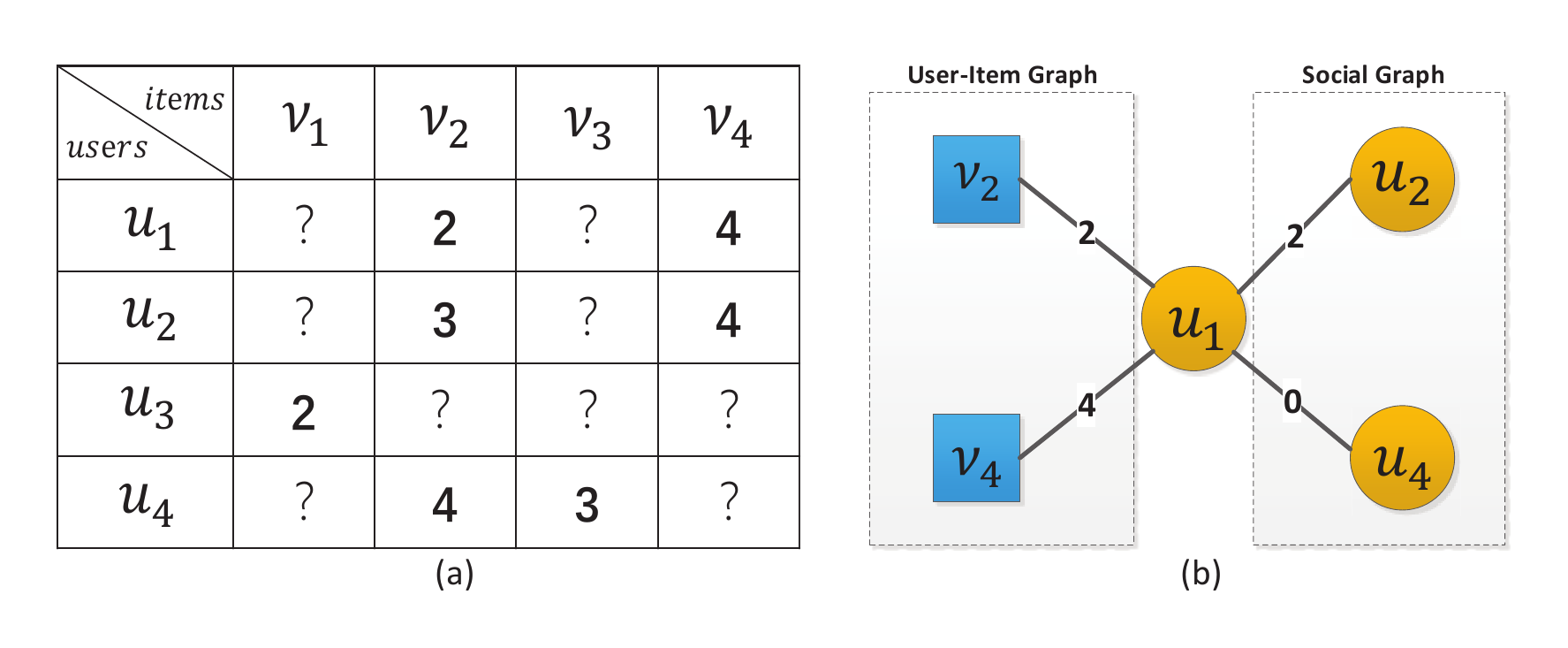}
	\caption{(a) The user-item rating matrix; (b) the interactions of the user $u_1$. }
	\label{fig1}
	\vspace{-0.2cm}
\end{figure}

\begin{figure}[t]
	\centering
	\includegraphics[width=0.5\textwidth]{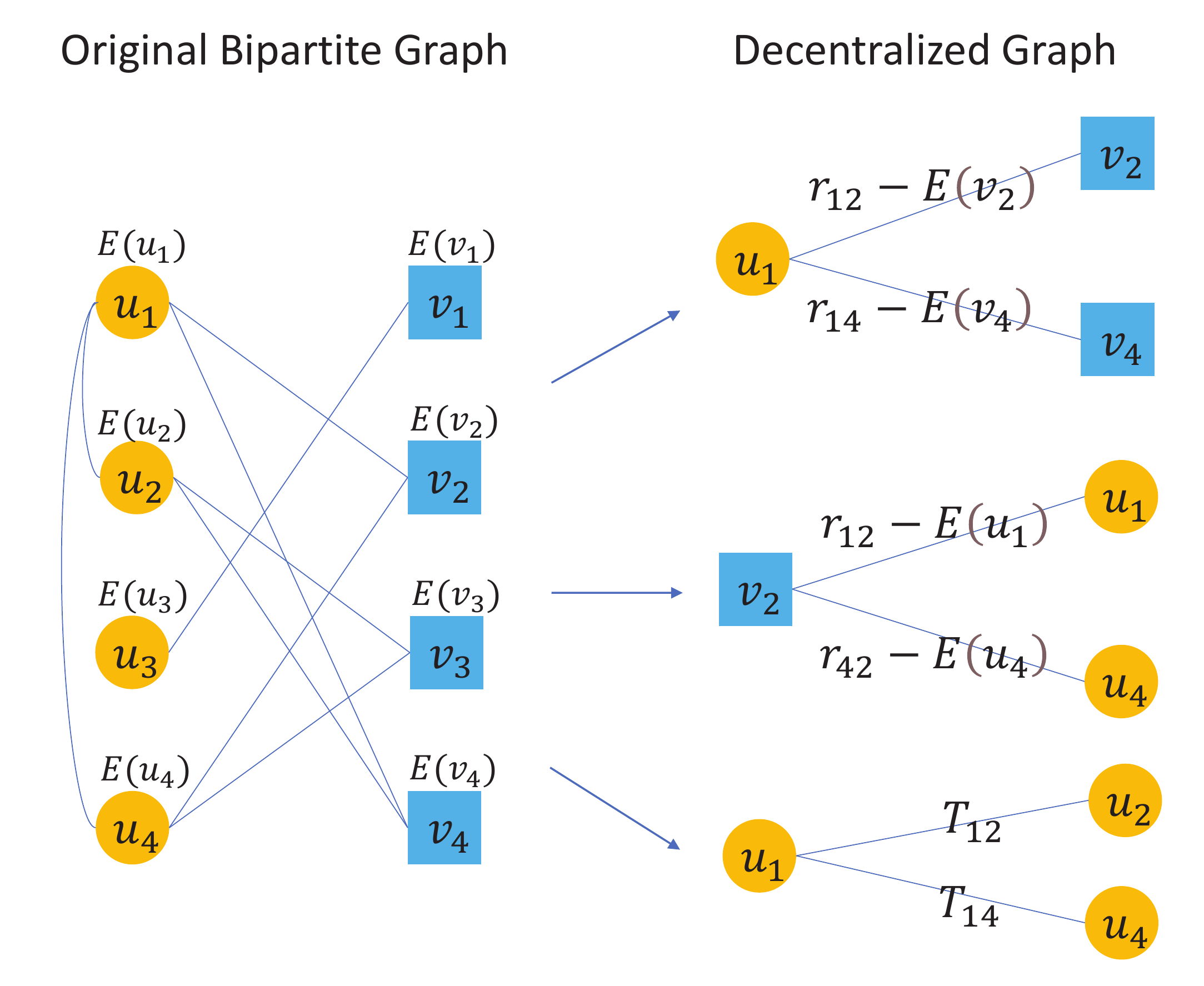}
	\caption{The original bipartite graph can be processed as a decentralized graph.}
	\label{fig2a}
\end{figure}

\begin{figure}[t]
\setlength{\abovecaptionskip}{-0.1cm} %缩小caption和图像之间的距离
	\centering
	\includegraphics[width=0.5\textwidth]{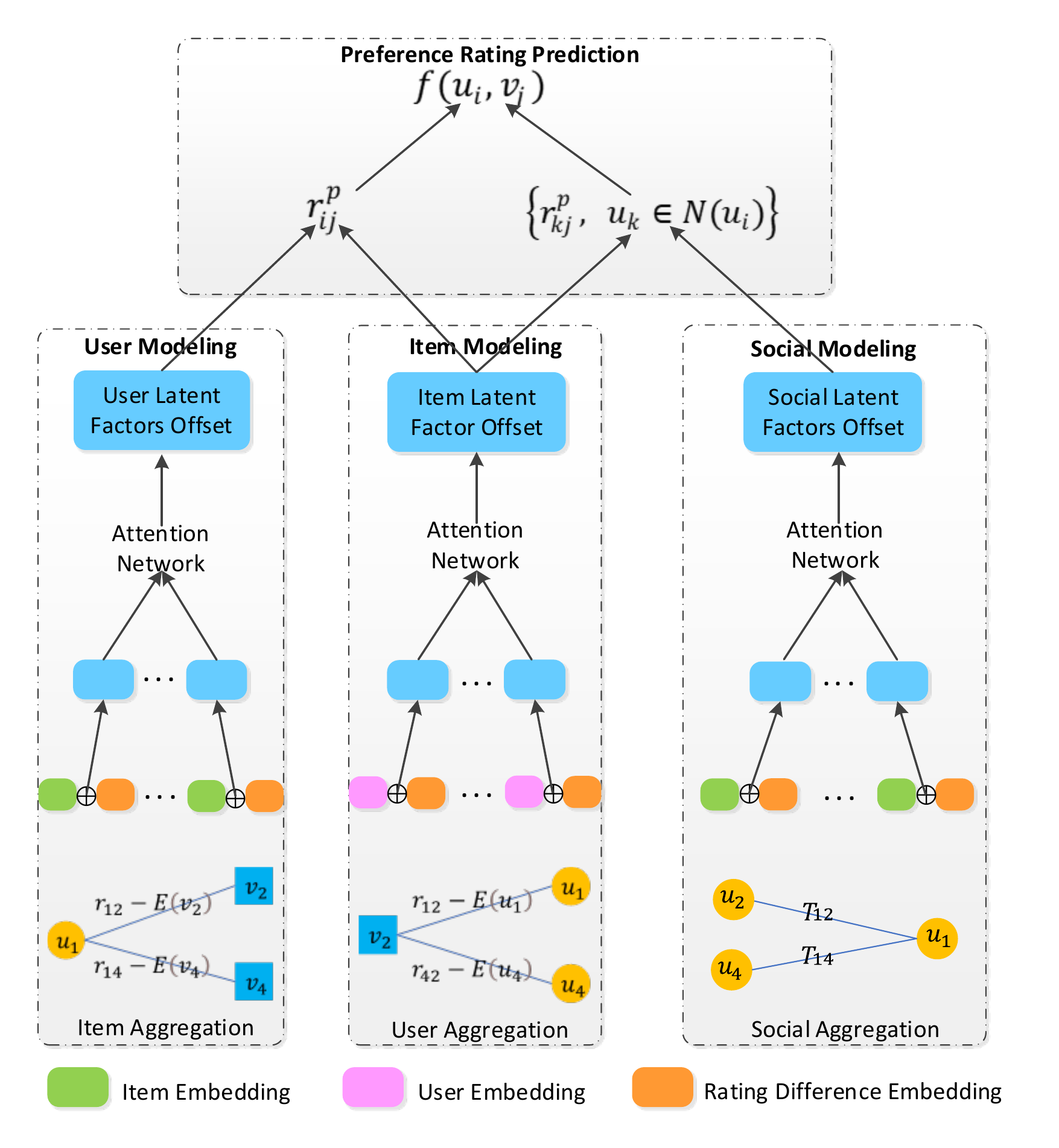}
	\caption{The overview of the proposed framework. It contains four components: user modeling, item modeling, social modeling and preference rating prediction.}
	\label{fig2}
\end{figure}

\subsection{Problem Formulation}
\label{subsecproblem}
Suppose that there are $N$ users $\mathcal{U} = \{u_1, u_2, \cdots, u_N\}$ and $M$ items $\mathcal{V} = \{v_1, v_2, \cdots, v_M\}$. As illustrated in Fig. \ref{fig1}(a), each user rates some items. If the rating can be observed, the rating score is greater than 0, otherwise there are missing values. %The higher the rating is, the more the user likes the item. 
The user-item rating matrix is denoted by $\mathbf{R} \in \mathbb{R}^{N \times M}$. %In this paper, we consider the task of rating prediction which 
The task is to predict the unobserved ratings in $\mathbf{R}$ and then return a ranked list of items for recommendation. To this end, the rating history and social relationships of users are employed to solve this problem. For example, we assume that the items $v_2$ and $v_4$ are rated by the user $u_1$ in Fig. \ref{fig1}(a), and the user $u_1$ has social relationships with the users $u_2$ and $u_4$ directly, as illustrated in Fig. \ref{fig1}(b). The left part of Fig. \ref{fig1}(b) containing the interactions between the users and the items is called user-item graph, and the right part containing the interactions between the users is called social graph. For the 
value in the edge of user-item graph, it represents the user $u_i$'s rating $r_{ij}$ on the item $v_j$. For the value in the social graph, we define it as the relationship coefficient between the users $u_i$ and $u_j$, i.e.,
\begin{equation}
   T_{ij} = 1 + \sum\limits_{v_k \in \{R(u_i)\cap R(u_j)\}} I\left(\left|r_{ik}-r_{jk}\right| \leq \delta \right), 
\end{equation}
where $I(x) = 1$ when $x$ is satisfied and zero otherwise, $\delta$ is the threshold for evaluating two users whether they like the same one item. 
%We do not consider using $I\left(\left|r_{ik}-r_{jk}\right|=0)$ for two reasons. One is that data with equal ratings between users and friends is very sparse, the other is that the ratings may not be the same for the same preference. To this end, the condition is slack. 
The relationship coefficient $T_{ij}$ represents the explicit relationship strength between the users, and denotes how much the user $u_i$ is similar to the socially connected user $u_j$. The higher the relationship coefficient is, there are more common items the two users like or dislike, in other words, the two users are more similar with each other. In order to predict the ratings of the user $u_1$ on the items $v_1$, $v_3$, we use the data described in Fig. \ref{fig1}(b) and the predicted ratings can be obtained by employing the GDSRec.

\subsection{General Framework}
Generally, the original data (e.g. Fig. \ref{fig1}) can be treated as a bipartite graph, as the left side in Fig. \ref{fig2a}, in which users have relations with other users and interactions with items. However, directly learning from such graphs may lead to the misunderstanding of the real user preference. Considering that a low rating from a fastidious user may not denote a negative attitude on this item because the user tends to assign low ratings to all items. Such phenomenon is not uncommon. If we only use the original data to learn the representations of users and items, this bias may lead to sub-optimal solution. To alleviate the above issue due to these user behaviors, we believe the statistical information can be utilized to address the bias offsets of users and items. Motivated by this idea, the original bipartite graph can be processed as a decentralized graph, as shown on the right side of Fig. \ref{fig2a}. For each user-item interaction, we subtract it from the centralized mean average. Then, the decentralized graph is utilized to train our model. We will give more details in the follow. 

In Fig. \ref{fig2}, we show an overview of the proposed model. This model includes four components: user modeling, item modeling, social modeling and preference rating prediction. For the user modeling, its aim is to learn the latent factor offsets of the users. The function of the social modeling is similar to that of the user modeling. The difference between two modelings is that the user modeling only models one user while the social modeling needs to integrate the learning for the social relations of the user simultaneously. The item modeling is used to learn the latent factors of the items.

As mentioned before, we solve the rating prediction problem by exploiting the decentralized graph data consisting of the decentralized user-item graph and the social graph. It is intuitive to obtain the final predicted rating $\hat{r}_{ij}$ between the user $u_i$ and the item $v_j$, including three components: the average rating $E(u_i)$ of the user $u_i$, the average rating $E(v_j)$ of the item $v_j$, and the final preference rating between the user $u_i$ and the item $v_j$, i.e.,
\begin{equation}
\label{301}
\hat{r}_{ij} = \frac{1}{2}[E(u_i) + E(v_j)] + f(u_i, v_j),
\end{equation}
where $E(u_i)$ and $E(v_j)$ set the benchmark for the prediction, $f(u_i, v_j)$ computes the final preference rating between the user $u_i$ and the item $v_j$. The function $f(u_i, v_j)$ can be expressed as
\begin{equation}
\label{302}
f(u_i, v_j) = \frac{1}{2}\left(r_{ij}^p + \sum\limits_{u_k \in N(u_i)} \lambda_{ik}r_{kj}^p\right),
\end{equation}
with
\begin{equation}
\label{303}
\lambda_{ik} = \frac{T_{ik}}{\sum\limits_{u_k \in N(u_i)}T_{ik}},
\end{equation}
where $r_{ij}^p$ is the preference rating between the user $u_i$ and the item $v_j$. The final predicted preference rating can be understood to consist of the user's own opinions and references of his socially connected users' ratings.
Since it is easy to obtain $E(u_i)$ and $E(v_j)$ from the original data, the key problem is how to derive the preference rating $r_{ij}^p$ between the user $u_i$ and the item $v_j$. To this end, we utilize the decentralized user-item graph to learn the representations of the users and the items. These representations are called latent factor offsets, because the proposed model maps the users and the items into a latent factor space by exploiting the decentralized graph data. In order to obtain the latent factor offsets of the users and the items, different data are employed to accomplish different goals. With the example in Fig. \ref{fig1}, the items interacted with the user (i.e., item aggregation of $v_2$ and $v_4$) are utilized for learning the latent factor offset of the user $u_1$. For the latent factor offset of item $v_4$, it is learned from the users whom the item interacts with (i.e., user aggregation of $u_1$ and $u_2$). The social offsets of $u_1$ is learned by performing social 
aggregation between the users that $u_1$ is socially connected with (i.e., $u_2$ and $u_4$). The preference rating prediction component is to learn the model parameters via prediction by integrating the user, item and social modeling components. It should be noted that if there is a new user or item without interaction records, the average rating of this user or item can be replaced by the global average value. The details of these model components are discussed as follows.

\subsection{User Modeling}
In this subsection, we detail how to model the latent factor offset (denoted as $\mathbf{h}_{u_i} \in \mathbb{R}^D$) of the user $u_i$ from item aggregation.

It can be seen that the decentralized user-item graph contains the interactions history between the users and the items, and the users' ratings on these items. In \cite{FanMa19}, the authors provided an approach to capture the interactions and the ratings for learning the latent factor of the user $u_i$. However, this approach does not reflect the statistical difference between $u_i$ and other users. As a result, instead of using the rating directly, we exploit the rating difference $\bar{r}_{ij}$ in the user modeling, i.e.,
\begin{equation}
\label{baruserrij}
\bar{r}_{ij} =  \lceil|r_{ij} - E(v_j)|\rceil.
\end{equation}
We create an embedding lookup table to map each difference $\bar{r}_{ij}$ into the table and one can easily obtain the difference representation $\mathbf{s}_{\bar{r}_{ij}}$ in this table. The reason that we do not use $r_{ij} - E(v_j)$ directly is that it is not convenient to use the embedding method in codes due to decimals. We believe quantitative methods can be utilized to tackle this problem and we leave it for future works.

To get the latent factor offset $\mathbf{h}_{u_i}$ for the user $u_i$ mathematically, we use the following function as
\begin{equation}
\mathbf{h}_{u_i} = \text{Tanh}\left(\mathbf{W} \cdot G_I(\{\mathbf{x}_{il}, \forall v_l \in R(u_i)\}) + \mathbf{b}\right),
\end{equation}
where $\mathbf{x}_{il}$ is the representation vector denoting the rating-difference-aware interaction between the user $u_i$ and the item $v_l$, $G_I$ is the item aggregation function, $\mathbf{W}$ and $\mathbf{b}$ are the weight and bias of a neural network, respectively. The purpose of the rating-difference-aware interaction is to capture the users' preference differences, which can help us to model the users' latent factor offsets. This is different from directly obtaining users' preferences in the past \cite{BergKipf17,FanMa19}. 
For the interaction between the user $u_i$ and the item $v_l$ with the rating difference $\bar{r}_{il}$, we model this interaction representation $\mathbf{x}_{il}$ as
\begin{equation}
\mathbf{x}_{il} = L_U([\mathbf{q}_{v_l} \oplus \mathbf{s}_{\bar{r}_{il}}]),
\end{equation}
where $L_U$ is a Multi-Layer Perceptron (MLP). As mentioned in the introduction, this method treats the bias as a vector and fuses it into the process of learning the user representation. In this way, we could better capture bias hidden in user interaction records.

Consider that each interaction between one user and the interacted items contributes differently to the user's latent factor offset. Inspired by the attention mechanisms \cite{YangYang16,ChenZhang18}, we define the item aggregation function $G_I$ as
\begin{equation}
G_I(\{\mathbf{x}_{il}, \forall v_l \in R(u_i)\}) = \sum\limits_{v_l \in R(u_i)}\eta_{il}\mathbf{x}_{il},
\end{equation}
where $\eta_{il}$ is the attention weight of the interaction between the user $u_i$ and the item $v_l$. In this way, the model can better capture the differences in the users' preferences. The core problem is how to get the attention weight. We take the following attention network to solve it.

\subsubsection*{Attention Network}
The input of this network is the interaction representation $\mathbf{x}_{il}$ and the user $u_i$'s embedding vector $\mathbf{p}_{u_i}$. According to \cite{FanMa19}, we exploit a two-layer neural network,
\begin{equation}
\dot{\eta}_{il} = \mathbf{w}_2^ T \cdot \text{ReLU}\left(\mathbf{W}_1 \cdot [\mathbf{x}_{il} \oplus \mathbf{p}_{u_i}] + \mathbf{b}_1 \right) + b_2,
\end{equation}
where $\text{ReLU}$ is rectified linear unit. The attention weight $\eta_{il}$ is obtained by normalizing above attentive scores with Softmax function, i.e.,
\begin{equation}
\label{atteta}
\eta_{il} = \frac{\exp(\dot{\eta}_{il})}{\sum\limits_{v_l \in R(u_i)} \exp(\dot{\eta}_{il})}.
\end{equation}
Finally, the latent factor offset $\mathbf{h}_{u_i}$ for the user $u_i$ can be written as
\begin{equation}
\mathbf{h}_{u_i} = \text{Tanh}(\mathbf{W} \cdot \left\{\sum\limits_{v_l \in R(u_i)} \eta_{il}\mathbf{x}_{il} \right\}+ \mathbf{b}).
\end{equation}

\subsection{Item Modeling}
This part aims to learn the latent factor offset $\mathbf{h}_{v_j}$ of the item $v_j$ from the user aggregation in the decentralized user-item graph. The user aggregation contains all users who interact with the item $v_j$, as well as users' ratings on $v_j$. Different users may express different attitudes towards the same item. This can help us to characterize the item to some extent. In order to describe the different characteristics of the item on different users, we modify the way of using the rating utilized in the user modeling. A new rating difference $\bar{r}_{ij}$ between the user $u_i$ and the item $v_j$  is defined as
\begin{equation}
\label{tildeitemrij}
\tilde{r}_{ij} = \lceil|r_{ij} - E(u_i)|\rceil.
\end{equation}
Exploiting this type of rating difference, we use the model to learn the latent factor offset of one item from different users.
The following whole process is similar to the user modeling.
For the interaction between the user $u_k$ and the item $v_j$ with the rating difference $\tilde{r}_{kj}$, we present a rating-difference-aware interaction representation $\mathbf{y}_{jk}$ composed by the user embedding $\mathbf{p}_{u_k}$ and the rating difference embedding $\mathbf{s}_{\tilde{r}_{kj}}$, i.e.,
\begin{equation}
\mathbf{y}_{jk} = L_I([\mathbf{p}_{u_k} \oplus \mathbf{s}_{\tilde{r}_{kj}}]),
\end{equation}
where $L_I$ is a MLP same as $L_U$ and the method for getting $\mathbf{s}_{\tilde{r}_{kj}}$ is the same as for $\mathbf{s}_{\bar{r}_{ij}}$ in the user modeling. For learning the latent factor offset $\mathbf{h}_{v_j}$, we introduce the function
\begin{equation}
\mathbf{h}_{v_j} = \text{Tanh}(\mathbf{W} \cdot G_U(\left\{\mathbf{y}_{jk}, \forall u_k \in R(v_j)\right\}) + \mathbf{b}),
\end{equation}
where $G_U$ is the user aggregation function. After introducing the attention mechanism for differentiating the contributions of users' interactions to $\mathbf{y}_{jk}$, we have
\begin{equation}
G_U(\left\{\mathbf{y}_{jk}, \forall u_k \in R(v_j)\right\}) = \sum\limits_{u_k \in R(v_j)} \xi_{jk}\mathbf{y}_{jk},
\end{equation}
where $\xi_{jk}$ is the attention weight obtained by using a two-layer neural attention network taking $\mathbf{y}_{jk}$ and $\mathbf{q}_{v_j}$ as the input. It can be written as
\begin{equation}
\label{attxi}
\xi_{jk} = \frac{\exp(\dot{\xi}_{jk})}{\sum\limits_{u_k \in R(v_j)}\exp(\dot{\xi}_{jk})},
\end{equation}
with
\begin{equation}
\dot{\xi}_{jk} = \mathbf{w}_2^T \cdot \text{ReLU}(\mathbf{W}_1 \cdot [\mathbf{y}_{jk} \oplus \mathbf{q}_{v_j}] + \mathbf{b}_1) + b_2.
\end{equation}
Similar to $\mathbf{h}_{u_i}$, $\mathbf{h}_{v_j}$ can be expressed as
\begin{equation}
\mathbf{h}_{v_j} = \text{Tanh}(\mathbf{W} \cdot \left\{\sum\limits_{u_k \in R(v_j)} \xi_{jk}\mathbf{y}_{jk}\right\} + \mathbf{b}).
\end{equation}

\subsection{Social Modeling}
The function of the social modeling is similar to that of the user modeling. When the user modeling learns the latent factor offset of the user $u_i$, this social modeling aims to learn the latent factor offsets of $u_i$'s socially connected users. For example, in Fig. \ref{fig2}, when the user modeling learns the latent factor offset of the user $u_1$, the social modeling learns the latent factor offsets of $u_2$ and $u_4$ in the way the user modeling does, respectively. 
Hence, we can directly obtain $\{\mathbf{h}_{u_k}, \forall u_k \in N(u_i)\}$. It should be noted that in this module, data is still the decentralized graph data. The main purpose of this module is to help target users calibrate their ratings through their social users. When calculating the rating between a user and an item, the preference of socially connected users of this user is an important reference. It can help the model get more accurate ratings.

\subsection{Rating Prediction}
After acquiring the latent factor offsets $\mathbf{h}_{u_i}$, $\mathbf{h}_{v_j}$ and $\{\mathbf{h}_{u_k}, u_k \in N(u_i)\}$ of the user $u_i$, the item $v_j$ and $u_i$'s social-connected users, respectively, the preference rating can be obtained using a three-layer neural network. For the preference rating $r^p_{ij}$, we use the following process to obtain it,
\begin{align}
\mathbf{z}_1 &= \text{Tanh}(\mathbf{W}_1 \cdot [\mathbf{h}_{u_i} \oplus \mathbf{h}_{v_j}] + \mathbf{b}_2),\\
\mathbf{z}_2 &= \text{Tanh}(\mathbf{W}_2 \cdot \mathbf{z}_1 + \mathbf{b}_2),\\
r^p_{ij} &= \mathbf{w}^T \cdot \mathbf{z}_2.
\end{align}
For $u_i$'s social users, their preference ratings $\{r^p_{kj}, \forall u_k \in N(u_i)\}$ are derived in the same way. Then using the expressions in \eqref{301}--\eqref{303}, we can obtain the final rating prediction between the user $u_i$ and the item $v_j$. It should be noted that in the testing stage, the average ratings of users and items are consistent with that in the training stage.

\subsection{Model Training}
We evaluate our proposed model from two perspectives including rating prediction and ranking prediction.  
%To train our model, two objective functions need to be established for above two tasks respectively. 
For the task of rating prediction, there is a commonly used objective function
\begin{equation}
\mathcal{L}_1 = \frac{1}{2\langle\mathcal{O}\rangle} \sum\limits_{(u_i, v_j) \in \mathcal{O}}(\hat{r}_{ij} - r_{ij})^2,
\end{equation}
where $r_{ij}$ is a ground truth rating rated by user $u_i$ on item $v_j$.

Learning to rank focuses on providing the end-user a ranked list of items \cite{RafailidisCrestanires17} and is widespread in different kinds of recommendation scenarios, e.g., top-k recommendation, sequential recommendation \cite{RendleChristophuai09,ChenXuwsdm18}.
% the performance at the top positions of the list, with the recommended items that end-users will actually see \cite{RafailidisCrestanires17}. Ranking problems are widespread in different kinds of recommendation tasks, e.g., top-k recommendation, sequential recommendation \cite{RendleChristophuai09,ChenXuwsdm18}.
In this task, for each user, the interacted items are labeled as 1 (i.e., positive samples) if the rating is equal to or greater than $F$, or 0 (i.e., negative samples) if not. This is to say, the users would like to click on or browse the items with a rating equal to or greater than $F$. As the two datasets in our experiment contain ratings from 1 to 5, we set up two cases including $F=3$ and $F=4$ in our experiments. The output prediction scores of all models is adjusted using the sigmoid function. For the ranking task, we choose the binary cross-entropy loss as the objective function
\begin{equation}
\mathcal{L}_2 = \sum\limits_{(u_i, v_j) \in \mathcal{O}} y_{ij}\log(\hat{y}_{ij}) + (1-y_{ij})\log(1-\hat{y}_{ij}),
\end{equation}
where $y_{ij}$ is a ground truth label of a sample and $\hat{y}_{ij}$ is a value between $(0,1)$ as predicted by the model. 

In the optimization of the objective functions, we adopt the RMSprop \cite{HintonSrivastava12} as the optimizer. It randomly selects a training instance, and updates each model parameter along the negative gradient direction. All the embedding vectors are initialized randomly and learned during the stage of training. For the rating difference embedding, it depends on the rating scale. In our experiment, each rating is in $\{1,2,3,4,5\}$. Hence one can set the input dimension of the embedding to be 5. 

To reduce the influence of overfitting and improve generalization performance, we apply the dropout strategy \cite{Srivastava14}. In particular, we introduce a node dropout strategy.

\subsubsection*{Node dropout}
In the decentralized user-item graph and the social graph, each user or item has a different number of interactions. For example, one user may have interacted with a dozen items, but another may have only interacted with a few items. In order to prevent the overfitting impact of too many interactions on the representation learning, we need to reduce some interactions in the training stage.
Based on the above ideas, node dropout is presented. In \cite{BergKipf17}, the authors introduced a way of dropping out nodes with a probability. However, this approach was detrimental to the nodes with few interaction records. Hence, in the process of learning the latent factor offsets for the users and the items, we reserve up to $K$ interaction nodes for each node randomly, such that we can protect the node where the learning resources are few.

In the next section, we validate the performance of the proposed model on two real-world datasets.

\vspace{-0.2cm}
\subsection{Time Complexity}
As described in the section \ref{subsecproblem}, there are $N$ users, $M$ items and $\langle\mathcal{O}\rangle$ training samples. We use node dropout and reserve up to $K$ interaction nodes for each node. At each module, we need to calculate two MLPs with small layers (e.g., two-layers). Thus the time cost for three modeling modules is about $O((N+M)KD)$. For the rating prediction module, the time cost is about $O(\langle\mathcal{O}\rangle D)$. Therefore, the total time cost is about $O(((N+M)K+\langle\mathcal{O}\rangle)D)$. In practice, as $K \ll \{N, M\}$, thus the total time complexity is acceptable.

%% file: experiments.tex
\vspace{-0.2cm}
\section{Experiment}
\label{sec4}
In this section, we conduct experiments to verify the effectiveness of the proposed GDSRec. We aim to answer the following research questions:

\textbf{RQ1:} How does GDSRec perform compared with existing methods, regarding both rating prediction and item ranking?
	
%\textbf{RQ2:} How does the negative sampling strategy affect the performance?

\textbf{RQ2:} How does the specific design of GDSRec affect the model performance? (i.e., ablation study)

\textbf{RQ3:} How does the node dropout strategy affect the model performance?

\textbf{RQ4:} How do the threshold $\delta$ affects the model performance?

% \vspace{-0.2cm}
\subsection{Dataset}
We evaluate our model on two benckmark datasets Ciao and Epinions\footnote{https://www.cse.msu.edu/\%7etangjili/trust.html}. They are taken from popular social networking websites. These two datasets contain users, items, ratings and social relations. The ratings are from 1 to 5. The statistics of Ciao and Epinions are shown in Table \ref{tabsta}.

\begin{center}
    \linespread{1.5}
    \begin{table}[]
    	\centering
    	\setlength{\belowcaptionskip}{0.2cm}
    	\caption{Statistics of the two datasets}\label{tabsta}
    	\rule{0pt}{15pt}
    	\begin{tabular}{c|c|c}
    		\hline  
    		Feature & Ciao & Epinions\\
    		\hline 
    		Users   & 7,317& 18,088\\
    		Items   & 10,4975& 261,649\\
    		Ratings & 283,319& 764,352\\
    % 		Density & 0.0368\%& 0.0161\%\\
    		\hline
    		Social Relations & 111,781 & 355,813\\
    % 		Density & 0.2087\% & 0.1087\%\\
    		\hline
    	\end{tabular}
    \end{table}
    % \vspace{-0.2cm}
\end{center}

\linespread{2}
\setlength{\tabcolsep}{1.1mm}{
% \usepackage{multirow}
% \usepackage[normalem]{ulem}
\begin{table*}[]
\setlength{\belowcaptionskip}{0.2cm}
\renewcommand{\arraystretch}{1}
\caption{Performance comparison of different recommender models}\label{tabres}
\begin{tabular}{|c|c|c|c|c|c|c|c|c|c|c|c|c|}
\hline
Training                        & Metrics & PMF    & FunkSVD & TrustMF & NeuMF  & DeepSoR & GCMC   & GCMC+SN & LightGCN & GraphRec     & Diffnet++    & GDSRec          \\ \hline
\multirow{2}{*}{Ciao(60\%)}     & MAE     & 0.9520 & 0.8462  & 0.7681  & 0.8251 & 0.7813  & 0.8157 & 0.7697  & 0.7715   & 0.7540       & \underline{0.7459} & \textbf{0.7328} \\ \cline{2-13} 
                                & RMSE    & 1.1967 & 1.0513  & 1.0543  & 1.0824 & 1.0437  & 1.0527 & 1.0225  & 1.0203   & 1.0093       & \underline{0.9987} & \textbf{0.9846} \\ \hline
\multirow{2}{*}{Ciao(80\%)}     & MAE     & 0.9021 & 0.8301  & 0.7690  & 0.8062 & 0.7739  & 0.8001 & 0.7526  & 0.7562   & \underline{0.7387} & 0.7398       & \textbf{0.7323} \\ \cline{2-13} 
                                & RMSE    & 1.1238 & 1.0515  & 1.0479  & 1.0617 & 1.0316  & 1.0323 & 0.9931  & 0.9963   & 0.9794       & \underline{0.9774} & \textbf{0.9740} \\ \hline
\multirow{2}{*}{Epinions(60\%)} & MAE     & 1.0211 & 0.9036  & 0.8550  & 0.9097 & 0.8520  & 0.8915 & 0.8602  & 0.8717   & 0.8441       & \underline{0.8435} & \textbf{0.8157} \\ \cline{2-13} 
                                & RMSE    & 1.2739 & 1.1431  & 1.1505  & 1.1645 & 1.1135  & 1.1258 & 1.1004  & 1.1103   & 1.0878       & \underline{1.0795} & \textbf{1.0685} \\ \hline
\multirow{2}{*}{Epinions(80\%)} & MAE     & 0.9952 & 0.8874  & 0.8410  & 0.9072 & 0.8383  & 0.8736 & 0.8590  & 0.8677   & \underline{0.8168} & 0.8201       & \textbf{0.8047} \\ \cline{2-13} 
                                & RMSE    & 1.2128 & 1.1422  & 1.1395  & 1.1476 & 1.0972  & 1.1052 & 1.0711  & 1.0801   & \underline{1.0631} & 1.0635       & \textbf{1.0566} \\ \hline
\end{tabular}
\end{table*}}

% \vspace{-1.5cm}
\subsection{Evaluation Metrics}
For evaluating the performance of the rating prediction of the proposed model, we adopt two well-known metrics, namely mean absolute error (MAE) and root mean square error (RMSE), to evaluate the predictive accuracy of the recommendation algorithms. The two metrics are defined as
\begin{align}
\text{MAE} = \frac{1}{\langle\mathcal{T}\rangle}\sum\limits_{u_i, v_j \in \mathcal{T}}|\hat{r}_{ij} - r_{ij}|,\\
\text{RMSE} = \sqrt{\frac{1}{\langle\mathcal{T}\rangle}\sum\limits_{u_i, v_j \in \mathcal{T}}(\hat{r}_{ij} - r_{ij})^2},
\end{align}
where $\mathcal{T}$ is the dataset of testing. Smaller values of MAE and RMSE indicate higher predictive accuracy.

In order to further validate the performance of the rating prediction, we extend it to item ranking evaluation. The testing set contains both positive (i.e., items whose score $\geq F$) and negative samples (items whose score $<F$). A ranked list of items in the testing set is assessed by using Recall and Normalized Discounted Cumulative Gain (NDCG) \cite{HeLiao17,WangHesigir19}. We compute Recall@5 by counting the number of positive samples contained in the top-5 positions of a ranked list. NDCG is a weighted version of Recall which assigns higher importance to the top positions in a list. Higher values of Recall and NDCG indicate better ranking performance. It should be noted that we only use the observed data to evaluate and do not rank all items in the item ranking evaluation. All the reported results are the average of five tests. 

% \vspace{-0.5cm}
\subsection{Baselines}
For comparison purposes, the following approaches are considered. We select four groups of representative methods including:
\begin{itemize}[leftmargin=*]
    \item Traditional recommendation algorithms: \textbf{PMF} \cite{MnihSalakhutdinov08}, it only uses rating information; \textbf{FunkSVD}\cite{koren2008factorization}, it considers user and item biases based on matrix factorization methods;
    \item Traditional social recommendation algorithm: \textbf{TrustMF} \cite{YangLei16}, which exploit social information on the basis of rating information;
    \item Deep neural 
    network based recommendation algorithms:
    \textbf{NeuMF} \cite{HeLiao17}, \textbf{DeepSoR} \cite{FanLi18}, which are classical deep and deep social recommendation, respectively.
    \item Graph neural network based recommendation algorithms:
    \textbf{GCMC}, \textbf{GCMC+SN} \cite{BergKipf17}, \textbf{GraphRec}\cite{FanMa19}, \textbf{LightGCN}\cite{he2020lightgcn} and \textbf{Diffnet++}\cite{wu2020diffnet++}.
\end{itemize}
%(1) traditional recommendation algorithms: \textbf{PMF} \cite{MnihSalakhutdinov08}, it only uses rating information; (2) traditional social recommendation algorithms: \textbf{SoRec} \cite{MaYang08}, \textbf{SoReg} \cite{MaZhou11}, \textbf{SocialMF} \cite{JamaliEster10} and \textbf{TrustMF} \cite{YangLei16}, these methods exploit social information on the basis of rating information; (3) deep neural network based recommendation algorithms: \textbf{NeuMF} \cite{HeLiao17}, \textbf{DeepSoR} \cite{FanLi18}, \textbf{GCMC+SN} \cite{BergKipf17}, \textbf{GraphRec} \cite{FanMa19}. 
% The baseline selection is the same with \cite{FanMa19}. Among the compared baselines, GraphRec is the state-of-the-art neural network-based social recommendation algorithm \cite{FanMa19}.

% \vspace{-0.5cm}
\subsection{Parameter Settings}
\label{paraset}
Our proposed model is implemented on the basis of Pytorch\footnote{http://pytorch.org/}. For two datasets, we select 80\% or 60\% as a training set to learn the parameters, and the rest are divided into a validation set and a testing set on average. This data split is consistent with \cite{FanMa19}. The validation set is used to tune hyper-parameters and the testing set is for the final performance comparisons. The threshold $\delta$ can be selected in $\{0,1,2,3\}$. We test the values of the embedding size $D$ in \{16, 32, 64, 128, 256, 512\}, and the interaction node reservation $K$ in \{5, 10, 15, 20\} on Ciao and \{15, 20, 25, 30\} on Epinions. The learning rate and the batch size are searched in \{$10^{-6}$, $10^{-5}$, $10^{-4}$, $5\times10^{-4}$\} and \{64, 128, 256\}, respectively. We stop the training if the sum of MAE and RMSE increases 10 successive epochs on the validation set. Model parameters and all the embedding vectors are initialized in default with a uniform distribution in Pytorch. For LightGCN\cite{he2020lightgcn} and Diffnet++\cite{wu2020diffnet++}, we adopt a setting of two-layer graph convolution. The hyper-parameters for these methods are tuned by grid search.

% \vspace{-0.5cm}
\subsection{Performance Comparison (RQ1)} 
% \vspace{-0.4cm}
\subsubsection{Rating prediction}
Table \ref{tabres} shows the performance comparison between different models for the task of rating prediction. Part of results of the compared methods are taken from \cite{FanMa19}.
%The rating prediction performance of the baseline algorithms are given in \cite{FanMa19} on Ciao and Epinions datasets, we show that in Table \ref{tabres}. The performance of our method GDSRec is given in the table at the same time. 
The values with underlines indicate the best performance between the baselines, to the best of our knowledge.
It can be seen from the table that FunkSVD outperforms PMF, which indicates that the user and item biases exist in practice. 
We can see that the traditional method TrustMF outperforms PMF and FunkSVD. When PMF and FunkSVD only use rating information, the traditional social recommender algorithm shows that the combination of rating and social information can effectively improve the recommendation performance. These results support that social network information can be leveraged when we solve the rating prediction problem using deep neural networks. 
 
NeuMF exploits neural network architecture to solve the problem, and it performs better than PMF. This shows the power of the neural network model in recommender algorithms. DeepSoR combines social information on the basis of the neural network, and performs better than NeuMF. In addition, comparing GCMC and GCMC+SN, the importance of social information also can be observed. 
LightGCN is the sate-of-the-art recommendation GNN-based model with implicit feedback, which models high-order user-item interaction but without exploiting social information. 
Generally, it outperforms GCMC but not as strong as GCMC+SN.
Both GraphRec and Diffnet++ take advantage of GNNs and combine them with the social network information for recommendations. These two approaches show that GNNs have good learning capabilities for representations. 
\begin{figure*}[]
\centering
    \includegraphics[width=18cm]{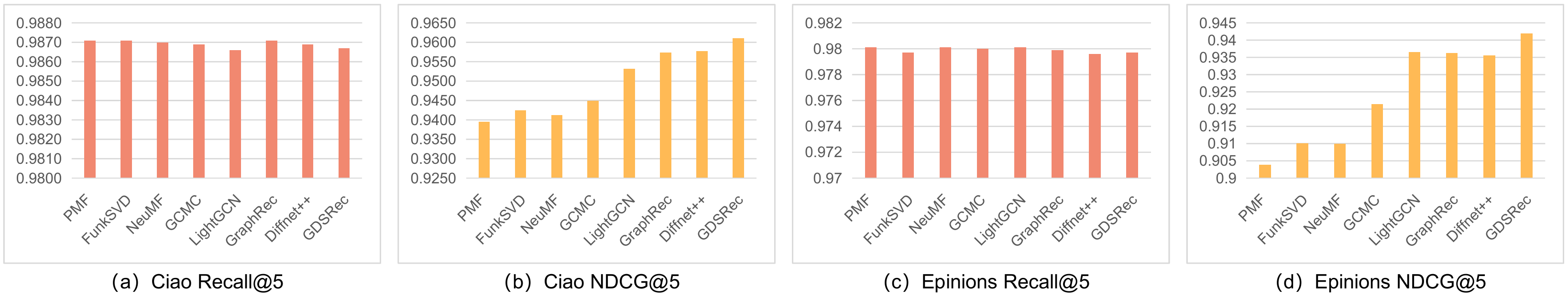}
	\caption{Performance of ranking on Ciao and Epinions datasets for $F=3$.}
	\label{figr6}
\end{figure*}
\begin{figure*}[]
\setlength{\abovecaptionskip}{-0.0cm} %缩小caption和图像之间的距离
	\centering
\includegraphics[width=18cm]{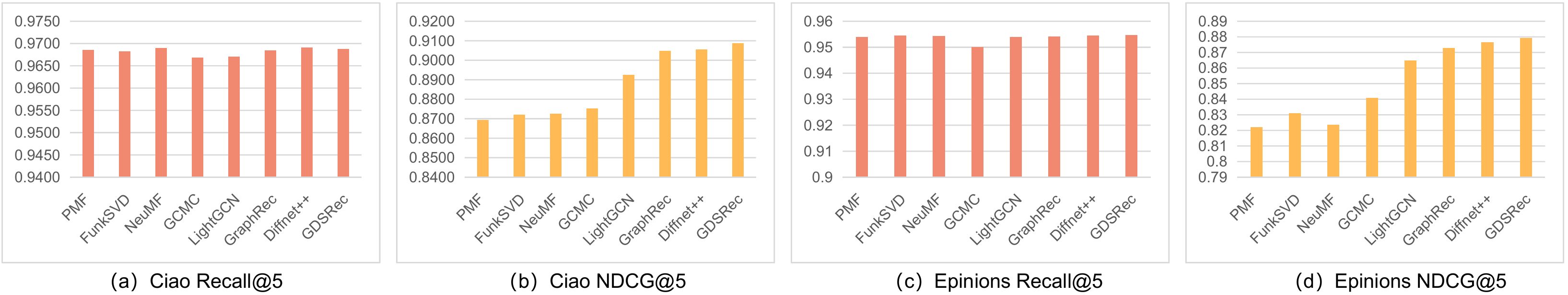}
	\caption{Performance of ranking on Ciao and Epinions datasets for $F=4$.}
	\label{figr7}
\end{figure*}

From Table \ref{tabres} we can see that our model GDSRec outperforms all other baseline methods. Compared to GraphRec and Diffnet++, our model exploits the users and items statistics, which helps to obtain the rating prediction benchmarks. The preference rating prediction is then sought by learning the latent factor offsets of the users and the items. In addition, unlike GraphRec and Diffnet++ which use the social network information to learn the user representations, our model uses the social network information as a method of correcting the final rating prediction. When the training set accounts for 60\% of the dataset, it can be seen that our model has an average performance improvement of 1.75\% over Diffnet++. When the training data is 80\% of the dataset, our model can obtain more performance improvements. Although the percentage of relative improvements are small, Koren has pointed out in \cite{Koren08} that even small improvements in MAE and RMSE may lead to significant differences of recommendations in practice.

% \vspace{-0.4cm}
\subsubsection{Item ranking}
In this part, we evaluate the performance of item ranking for the proposed model. For comparison purpose, we select the traditional algorithm PMF and FunkSVD, the classical deep algorithm NeuMF, GNN-based models including LightGCN, GraphRec and Diffnet++ to compare with our GDSRec. 
The results are shown in Fig. \ref{figr6} and \ref{figr7}. From the figures we can see that the four models show similar and high performance on the Recall evaluation metric. This is because the positive label takes up a large proportion in the two datasets. The models' predictions are naturally biased and give positive results with high probabilities for negative samples. This makes it difficult to judge the performance of the models on the Recall. In contrast, the performance of the models on the NDCG is different. This demonstrates a difference in the ranking ability of the four models. In sum, the GDSRec performs better than other counterparts in terms of NDCG. It further demonstrates that our model is more effective to push the positive items to high ranking positions.%. Note that the GraphRec and GDSRec both exploit the explicit feedback (i.e., ratings) to learning the representation of users and items. It illustrates the importance of users having explicit preferences for recommender systems.

% \vspace{-0.5cm}
\subsection{Ablation Study (RQ2)}
In this subsection, we further investigate the impact of the model components of the proposed GDSRec.
\begin{figure*}[t]
\setlength{\abovecaptionskip}{-0.00cm} %缩小caption和图像之间的距离
	\centering
	\includegraphics[width=18cm]{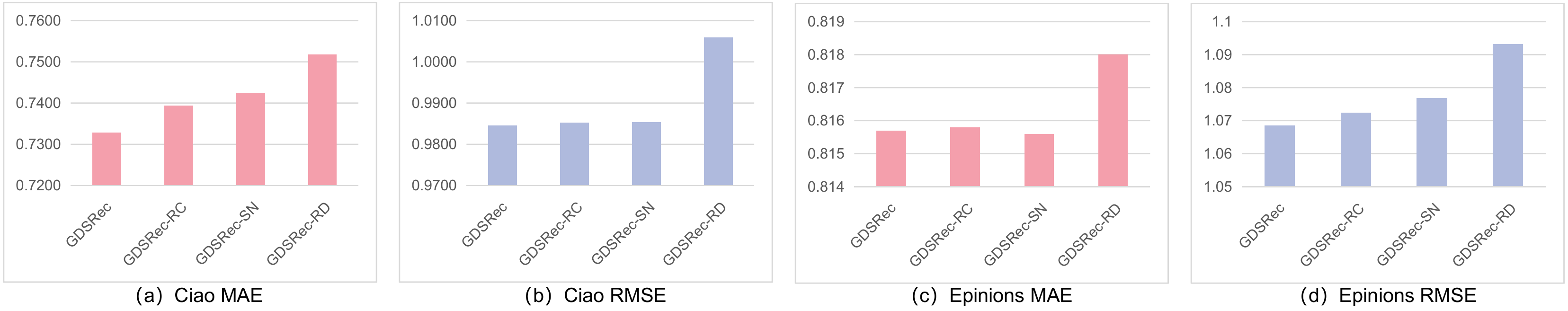}
	\caption{Effect of social network and user ratings on Ciao and Epinions datasets.}
	\label{fig3}
\end{figure*}

\begin{figure}[!h]
\centering
\includegraphics[width=7.5cm]{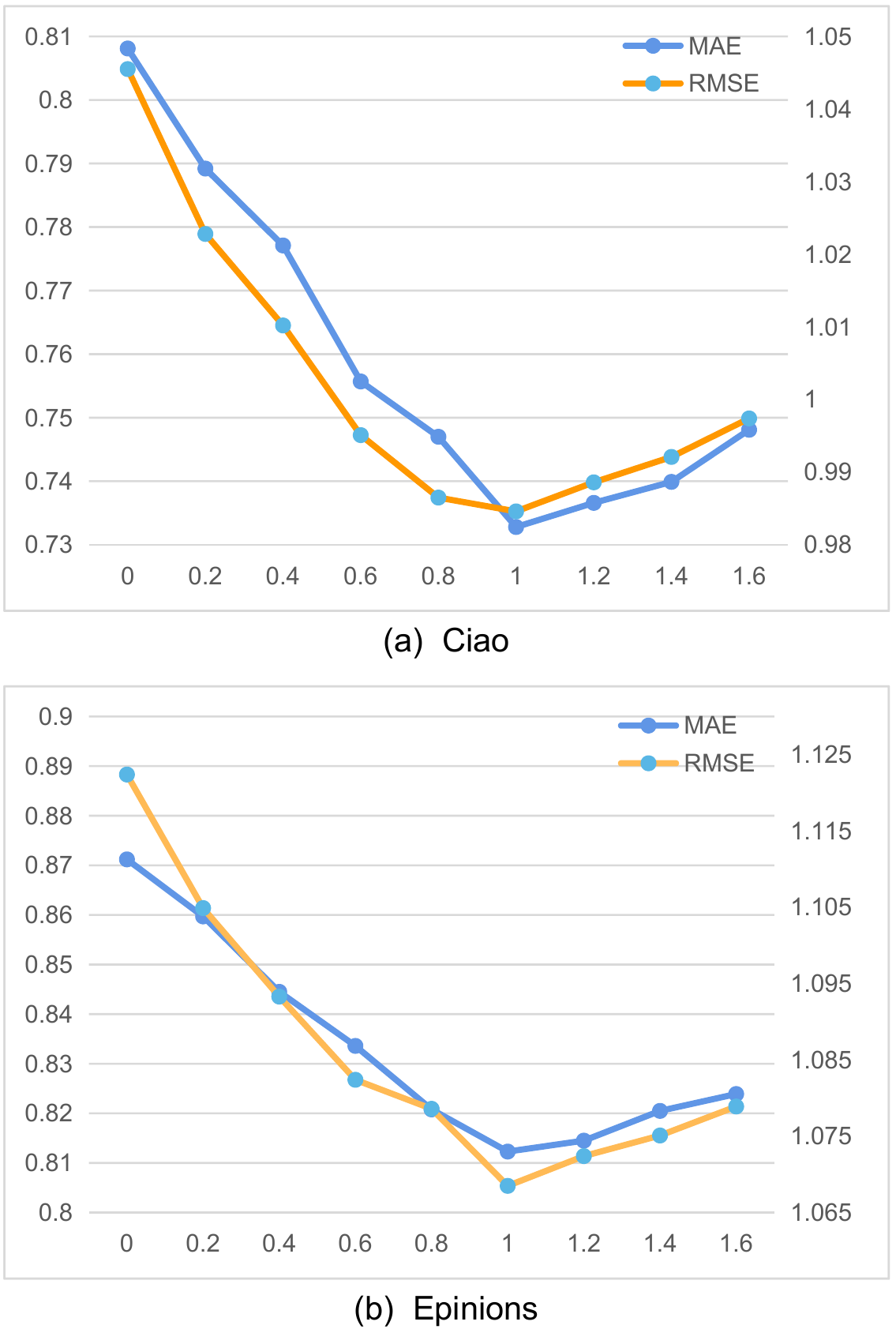}
\caption{Effect of $\alpha$ on Ciao and Epinions datasets.}
\label{fig10}
\end{figure}
% \vspace{-0.5cm}

\begin{figure*}[t]
\setlength{\abovecaptionskip}{-0.00cm} %缩小caption和图像之间的距离
	\centering
\includegraphics[width=18cm]{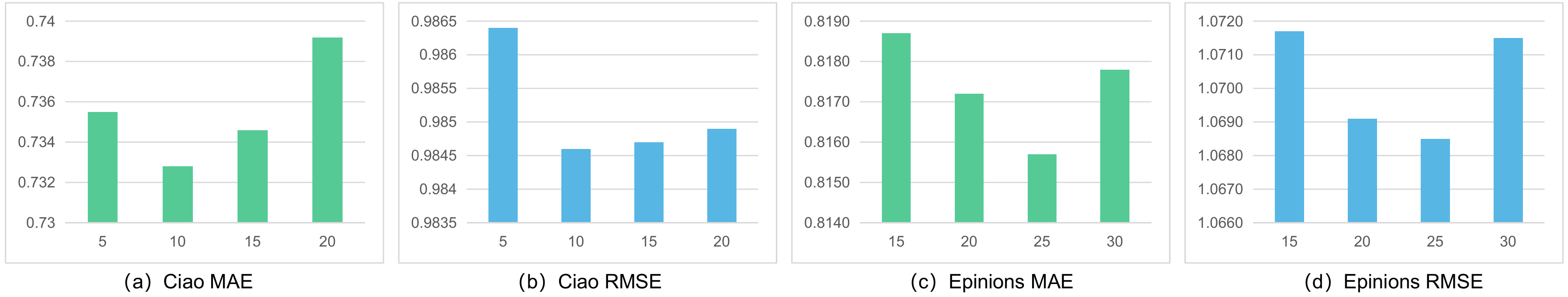}
	\caption{Effect of node dropout on Ciao and Epinions datasets.}
	\label{fig5}
\end{figure*}
% \vspace{-0.5cm}

% \vspace{-0.5cm}
\subsubsection{Effect of Social Network and User Ratings}
The effectiveness of the proposed model is presented in this subsection. Our model incorporates three factors: 1) the addition of the relationship coefficient for the social relations; 2) using the social relations to adjust the rating prediction; 3) learning the latent factor offsets by employing the statistics of the users and the items. To better understand the proposed model, we make several variants of the model and compare the performance between them. These variants are defined as:
\begin{itemize}[leftmargin=*]
	
	\item GDSRec-RC: The relationship coefficients of the social relations are removed from the proposed framework. This variant implies that all social relationships are equal and undifferentiated. In other words, all relationship coefficients are 1.
	
	\item GDSRec-SN: The social network information of the GDSRec is removed. This variant ignores the preference rating prediction of the social relations, and only uses $\mathbf{h}_{u_i}$ and $\mathbf{h}_{v_j}$ to obtain $f(u_i, v_j)$, i.e., $f(u_i, v_j) = r^p_{ij}$.
	
	\item GDSRec-RD: The latent factor offsets are learned from the rating difference in our proposed framework. This variant means that the latent factor offsets are learned with original rating data, rather than the rating difference. In other words, the rating difference ($\bar r_{ij}$, $\tilde r_{ij}$) defined in \eqref{baruserrij} and \eqref{tildeitemrij} is replaced by $r_{ij}$. 
\end{itemize}
For simplicity, we conclude these variants in Table \ref{variantsofGDSRec}. In the following, we compare the performance of these variants with that of the GDSRec.

\begin{center}
\linespread{1.5}
    \begin{table}[h]
        \centering
        \setlength{\belowcaptionskip}{0.2cm}
        \linespread{2}
        \caption{Modification of GDSRec}\label{variantsofGDSRec}
        \begin{tabular}{cc}
            \hline
            \multicolumn{2}{c}{Modification}                        \\ \hline
            \multicolumn{1}{c|}{GDSRec-RC} & \multicolumn{1}{c}{Eq. (\ref{303}) $\Rightarrow \lambda_{ik} = \frac{1}{\langle N(u_i) \rangle}$ } \\ \hline
            \multicolumn{1}{c|}{GDSRec-SN} & \multicolumn{1}{c}{Eq. (\ref{302}) $\Rightarrow f(u_i, v_j) = r^p_{ij} $ } \\ \hline
            \multicolumn{1}{c|}{GDSRec-RD} & \multicolumn{1}{c}{Eq. (\ref{baruserrij}) and (\ref{tildeitemrij}) $\Rightarrow  \bar{r}_{ij}, \tilde{r}_{ij} = r_{ij}$ } \\ \hline
        \end{tabular}
    \end{table}
\end{center}

The performance comparisons among the GDSRec and three variants regarding rating prediction are given in Fig. \ref{fig3}. The hyper-parameters in these models are set to be the same. From the results shown in the figure, we can conclude that:
\begin{itemize}[leftmargin=*]
	\item Impact of the Social Network: We now analyze the impact of the social network on the recommendation performance. First of all, we can see that the performance of the GDSRec-RC is slightly weaker than that of the GDSRec from the Fig. \ref{fig3}. Although the RMSE of the GDSRec-RC is similar to that of the GDSRec on Ciao, the MAE is 0.9\% higher than that of the GDSRec. For the dataset Epinions, it can be observed that while the MAE of the GDSRec-RC is close to that of the GDSRec, the RMSE of the GDSRec-RC is higher than that of the GDSRec. This verifies that the relationship coefficient is beneficial for the model. On the other hand, the RMSE of the GDSRec-SN is 0.79\% higher than that of the GDSRec while the MAE results of two models are close. This is to say, the social network is important for the recommendation performance.
	\item Impact of the Rating Difference: When we use the original rating data instead of the rating difference for training the latent factor offsets, we can see from Fig. \ref{fig3} that both MAE and RMSE of the GDSRec-RD on two datasets are much higher than that of the GDSRec. So the performance of the GDSRec-RD is much worse than that of the GDSRec. It validates that our core idea, processing on the original rating data, is very efficient, which helps to improve the performance of recommendation. We believe it can be applied directly to other models and lead to performance improvements, and we leave it for future works.
\end{itemize}

\linespread{1.5}
\rule{0pt}{15pt}
\begin{table}[h]
\centering
    \setlength{\belowcaptionskip}{0.2cm}
	\caption{Effect of attention network on Ciao and Epinions datasets.}\label{tabatt}
	\begin{tabular}{|c|c|c|c|c|}
		\hline
		Training&Metrics&GDSRec-avg&GDSRec-max&GDSRec\\
		\hline
		{Ciao(60\%)} & MAE &0.7326 &0.7388 &0.7328 \\
		\cline{2-5}
		&RMSE &0.9871 &0.9884 &0.9846 \\

		\hline
		{Epinions(60\%)} & MAE &0.8155 &0.8183 &0.8157 \\
		\cline{2-5}
		&RMSE &1.0704 &1.0706 &1.0685 \\
		\cline{2-5}
		\hline
	\end{tabular}
\end{table}

% \vspace{-0.8cm}
\subsubsection{Effect of Attention Network}
In this subsection, the effectiveness of the attention network is evaluated. The proposed GDSRec use softmax to normalize the attention scores. %In our design, the attention weights are obtained by normalizing the attentive scores with Softmax function. 
Here, we design two variants of the attention weights. One variant is to rewrite \eqref{atteta} and \eqref{attxi} as
\begin{equation}
\begin{cases}
	\eta_{il} = \frac{1}{\langle R(u_i)\rangle},\\
	\xi_{jk} = \frac{1}{\langle R(v_j) \rangle},
\end{cases}
\end{equation}
respectively. The other is,
\begin{equation}
\label{attmax}
\begin{cases}
	\eta_{il} = \max\limits_{v_l \in R(u_i)}\left(\frac{\exp(\dot{\eta}_{il})}{\sum\limits_{v_l \in R(u_i)} \exp(\dot{\eta}_{il})}\right),\\
	\xi_{jk} = \max\limits_{u_k \in R(v_j)}\left(\frac{\exp(\dot{\xi}_{jk})}{\sum\limits_{u_k \in R(v_j)}\exp(\dot{\xi}_{jk})}\right).
\end{cases}
\end{equation}
We use GDSRec-avg and GDSRec-max to denote these two variants respectively. Note that there is no another variant similar to the GDSRec-max which replaces maximum of \eqref{attmax} with minimum. This is because it may result in the output of the attention network very small if there is a attention weight is close to zero.

The performance comparison regarding rating prediction is shown in Table \ref{tabatt}. We can observe that the GDSRec-avg achieves the same performance with the GDSRec on the MAE. The performance of the GDSRec-max is always the worst. We can put the GDSRec-avg and GDSRec-max in the same category as these two variants both have the same output weight for different input attention scores. We believe that the GDSRec-avg performs better than the GDSRec-max as the adjustment of the GDSRec-avg is more gentle. In summary, the GDSRec assigns different weights to users or items when learning the representations and therefore has better performance.

% \vspace{-0.3cm}
\subsubsection{Effect of Average Rating}
According to \eqref{301}, the final prediction depends on the average ratings (i.e., $E(u_i)$ and $E(v_j)$). Here we discuss what the performance will be if the average ratings are changed. Toward this end, we change the weight of the average ratings in the final prediction, the expression \eqref{301} is rewritten as
\begin{equation}
\label{401}
\hat{r}_{ij} =  \frac{\alpha}{2}[E(u_i) + E(v_j)] + f(u_i, v_j),    
\end{equation}
where $\alpha$ is a hyper-parameter. 

We test the values of $\alpha$ in \{0, 0.2, 0.4, 0.6, 0.8, 1, 1.2, 1.4, 1.6\} and give the results in Fig. \ref{fig10}. From the figure we can see that the performance degrades significantly as $\alpha$ decreases from 1 to 0. It's easy to explain that the model can be thought of as fitting the decentralized training data. Thus, once the average ratings changed, the final prediction loses this information directly. It inevitably leads to performance degradation. As $\alpha$ goes large, the performance relatively deteriorates. These results indicate the importance of users' personalized preferences in predicting.

% \vspace{-0.5cm}
\subsection{Effect of Node Dropout (RQ3)}
Now, we analyze the effect of the node dropout on the performance of recommendation. The node dropout is used to avoid overfitting problems caused by too many interactions for nodes. For a limited number $K$ of interaction nodes, we show the results in Fig. \ref{fig5}, where the training set accounts for 60\% of the dataset and $D=256$. On the dataset Ciao, the GDSRec achieves the best performance at $K=10$ when $K$ increases from 5 to 20. For the dataset Epinions, the results are somewhat different. As $K$ goes from 15 to 30, we can clearly see that both MAE and RMSE are minimal at $K=25$. It verifies that the interaction nodes number $K$ effects the performance of the proposed model. For a new dataset, a limited number of interaction nodes $K$ needs to be tested experimentally.

\linespread{1.5}
\begin{table}[]
\centering
\setlength{\belowcaptionskip}{0.2cm}
\caption{Effect of the threshold $\delta$ on Ciao and Epinions datasets.}\label{delta}
        \begin{tabular}{|c|c|c|c|c|c|}
        \hline
        Training                        & Metrics & $\delta=0$      & $\delta=1$      & $\delta=2$      & $\delta=3$      \\ \hline
        \multirow{2}{*}{Ciao(60\%)}     & MAE     & 0.7327 & 0.7328 & 0.7441 & 0.7446 \\ \cline{2-6} 
                                        & RMSE    & 0.9874 & 0.9846 & 0.9856 & 0.9879 \\ \hline
        \multirow{2}{*}{Epinions(60\%)} & MAE     & 0.8242 & 0.8157 & 0.8252 & 0.8271 \\ \cline{2-6} 
                                        & RMSE    & 1.0709 & 1.0685 & 1.0696 & 1.0726 \\ \hline
        \end{tabular}
\end{table}

% \vspace{-0.5cm}
\subsection{Effect of the Threshold $\delta$ (RQ4)}
In Table \ref{delta}, we give the results of the MAE and RMSE on two datasets for different $\delta$ thresholds. Generally, when $\delta$ is 1, the performance on Ciao and Epinions is the best. As $\delta=0$, the data of the social relation is very sparse. Therefore, it can not help the model to learn better. And as $\delta$ becomes larger than 1, it may introduce some noise into the social relation data and causes poor performance.

%% file: conclusions.tex
\vspace{-0.5cm}
\section{Conclusions}
\label{sec5}
In this paper, we have proposed a novel framework GDSRec for the rating prediction problem in social recommendations. GDSRec treats rating biases as vectors and fuses them into the process of learning user and item representations. To the end, we have dealt with the original graph into the decentralized graph by utilizing the statistical information, and extracted bias information explicitly on the graph. It provides a decentralized perspective on learning the latent factor offsets for the users and the items. And the statistical information provides important benchmarks for the rating predictions. In addition, we have differentiated the explicit strengths of social relations for the users and added these strengths to the final predictions. Experiments on two real-world datasets have been conducted. The results have shown that our new method has better rating prediction performance than its counterparts. In addition, we have conducted experiments to verify the ability of ranking for the proposed model. To conclude, the proposed model achieves better performance on both rating prediction and item ranking.